\documentclass[prl,twocolumn,nofootinbib]{revtex4-2}
\usepackage[utf8]{inputenc}
\usepackage{graphicx}
\usepackage{amsmath}
\usepackage{cancel}
\usepackage[normalem]{ulem}
\usepackage{url}
\usepackage{hyperref}
\usepackage{color}
\usepackage{tikz}  % to show the diagram
\usepackage{lineno}  % to show line numbers
\usepackage{draftcopy}  % to show line numbers
\graphicspath{{./figs/}}

\newcommand{\kw}[1]{{\color{red}[\textbf{KW:}\,{#1}]}}

\begin{document}   

\title{Heavy flavor as a probe of hot QCD matter produced in proton-proton collisions}
\author{Jiaxing Zhao, Joerg Aichelin, Pol Bernard Gossiaux, and Klaus Werner}
\affiliation{SUBATECH, Nantes University, IMT Atlantique, IN2P3/CNRS, 4 rue Alfred Kastler, 44307 Nantes cedex 3, France}

\date{\today}

\begin{abstract}
The creation of a quark-gluon plasma (QGP) is expected in heavy ion collisions. It came as a surprise that proton-proton collisions at ultrarelativistic energies show as well  a ``QGP-like'' behavior and signs of the creation of a fluid, although the corresponding system size is not more than a few cubic femtometers. Even more surprisingly,  also heavy flavor particles seem to be part of the fluid or at least interact with it. In this paper, we will investigate in a quantitative way this ``collective behavior'' of heavy flavor, by employing the newly developed EPOS4HQ approach, which has proven to be compatible with basic experimental data of light flavor hadrons. We will investigate all observables, which may manifest collectivity, as particle spectra, elliptic flow, baryon-to-meson ratios, and two-particle correlations, and compare the results with experimental data. We will try to disentangle initial state effects, those being due to interactions between charm quarks and plasma partons, and final state effects (hadronization).
\end{abstract}
\maketitle 

%==================================
\section{Introduction}
There exists ample evidence that a new, deconfined phase of matter, predicted by Lattice QCD calculations ~\cite{Bernard:2004je,Bazavov:2011nk} and called quark-gluon plasma (QGP), is created in 
relativistic heavy-ion collisions.
The QGP expands rapidly leading to a continuous decrease in temperature and density. Finally, the energy density falls below a critical value and the QGP hadronizes into hadrons, which are finally observed.
Heavy mesons, which contain either a heavy quark $Q$ or a heavy antiquark $\bar Q$, have turned out to be an ideal probe to study the QGP due to two reasons: a) The heavy quark mass is much larger than the QCD cutoff, $m_Q\gg \Lambda_{QCD}$. Therefore their production can be well described by perturbative QCD (pQCD) and hence their initial momentum distribution is known;  
b) The heavy quark mass is much larger than the typical temperature of the hot medium, $m_Q\gg T_{QGP}$. Their masses are unchanged in the hot medium and their number is conserved during the evolution. 
The transverse momentum spectrum of heavy mesons in heavy-ion collisions (as compared to that in proton-proton collisions) can only be understood if a QGP is created in these collisions. This is confirmed by another key observable, the elliptic flow, $v_2$, which heavy quarks can only acquire through interactions with the QGP because initially they are formed in hard processes, which are azimuthally isotropic.

It came as a surprise that in (high multiplicity) $pp$ collisions observations have been made, which are considered in heavy-ion collisions as a proof of the existence of a QGP. They include the observation of long-range correlations (also called near-side ``ridge'')~\cite{CMS:2010ifv,ATLAS:2015hzw,CMS:2015fgy,CMS:2016fnw}, of strangeness enhancement~\cite{ALICE:2016fzo}, and of a finite elliptic flow of $D^0$ mesons~\cite{CMS:2020qul}. It was even more astonishing that the multiplicity of charmed baryons at mid-rapidity in these collisions ~\cite{ALICE:2018pal}, is considerably higher than expected from the analysis of $e^+e^-$ collisions, questioning the process independence of fragmentation functions~\cite{ALICE:2020wfu,ALICE:2021psx,CMS:2023frs,ALICE:2023wbx}.  This is also  the topic of several theoretical studies ~\cite{He:2019tik,Minissale:2020bif,Li:2021nhq,Beraudo:2023nlq} which assume that the hadronization mechanism of heavy quarks in high energy $pp$ collisions is quite different from that in $e^+e^-$.

The purpose of this letter is to show that the recently advanced EPOS4HQ\footnote{based on version EPOS4.0.1.s9} approach, which allows for the creation of a QGP in high energy density regions, independent of the system size, describes $pp$ collisions as well and can reproduce quantitatively the experimental observations. Therefore $pp$ collisions at ultrarelativistic energies are just the small system size limit of $AA$ collisions. 
%-----------------------------------------------
\section{EPOS4 primary interactions}

In the EPOS4 approach, we distinguish ``primary interactions'' and
``secondary interactions''. The former refer to parallel partonic
scatterings, happening at very high energies instantaneously, at $t=0$.
Any notion of a sequential ordering makes no sense. The theoretical
tool is S-matrix theory, using a particular form of the proton-proton
scattering S-matrix. The main new development in EPOS4 \cite{werner:2023-epos4-overview,werner:2023-epos4-heavy,werner:2023-epos4-smatrix,werner:2023-epos4-micro}
is a way to accomodate simultaneously: (1) rigorous parallel scattering,
(2) energy-momentum sharing, and (3) validity of the AGK theorem \cite{Abramovsky:1973fm},
which assures binary scaling (in $AA$ scattering) and factorization
\cite{Collins:1989} (in $pp$) for hard processes, by introducing
(in a very particular way) saturation, compatible with recent ``low-x-physics''
considerations \cite{Gribov:1983ivg,McLerran:1993ka,kov95,kov96,jal97,jal97a,jal99a}.
So although energy-momentum sharing makes things complicated, it is
not only mandatory for a consistent picture, it also allows to understand
a connection between factorization and saturation.

\begin{figure}[h]
\centering{}\includegraphics[scale=0.22]{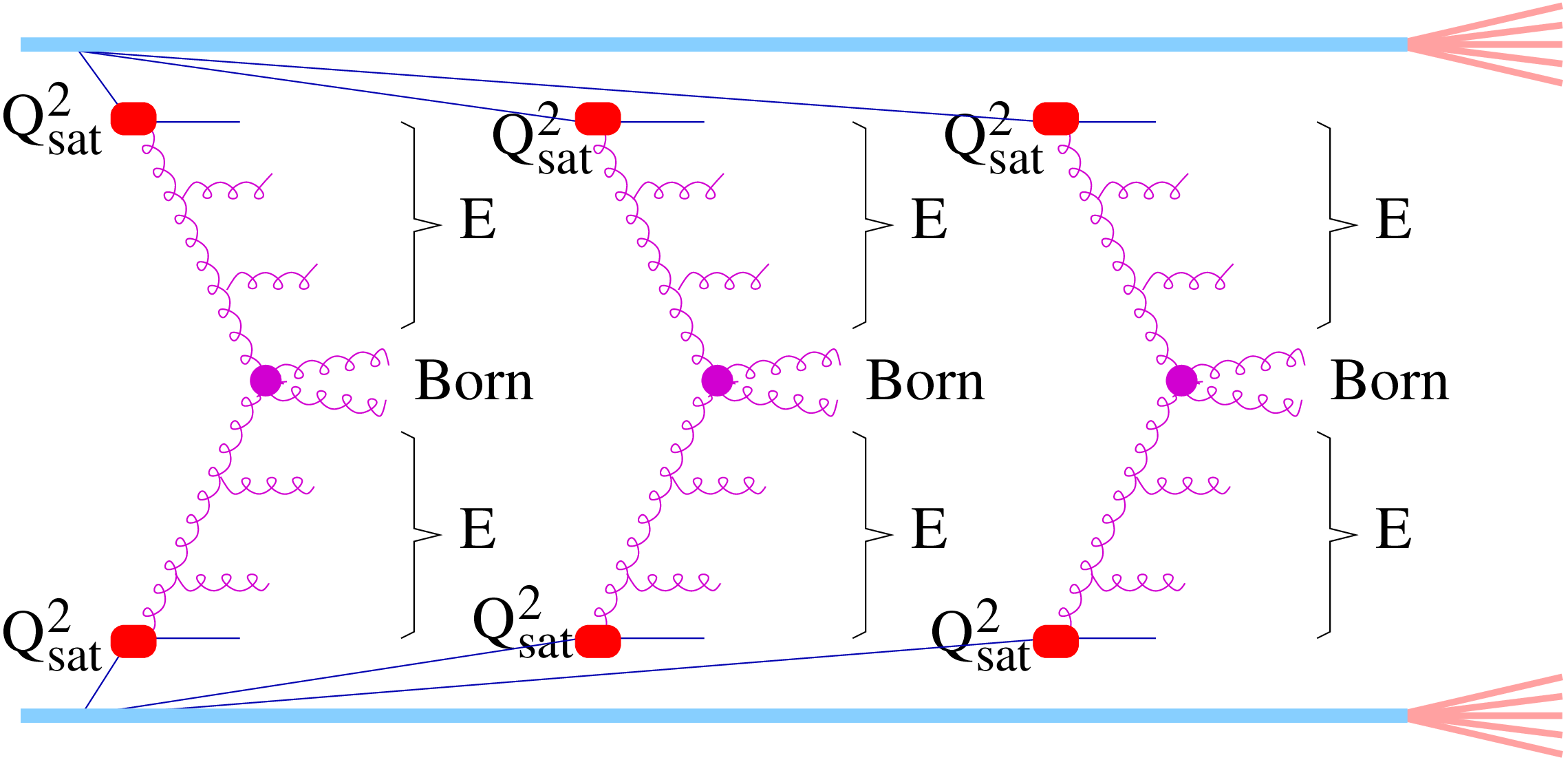} \caption{Rigorous parallel scattering scenario, here for three parallel scatterings.
The red symbols should remind us that the parts of the diagram representing
nonlinear effects (like gluon fusion) are replaced by simply using
saturation scales. \label{saturation-three-pom}}
\end{figure}

Validity of AGK means that we can do the same as models based on factorization
(defining and using parton distribution functions) to study hard processes
(this is needed to prove consistency), but we can do much more. One
of the highlights of the past decade in our domain, concerns collective
phenomena in small systems. It has been shown that high-multiplicity
pp events show very similar collective features as earlier observed
in heavy ion collisions \cite{CMS:2010ifv}. High multiplicity means
automatically ``multiple parton scattering'', and the EPOS4 formalism
allows perfectly to treat this. Here one does not employ the usual
parton distribution functions (representing the partonic structure
of a fast nucleon), one treats the different scatterings (happening
in parallel) individually, for each one a parton evolution is realized
according to some evolution function $E$ (representing the space-like
(SL) cascascade), as sketched in Fig. \ref{saturation-three-pom}.
One still has DGLAP evolution \cite{GribovLipatov:1972,AltarelliParisi:1977,Dokshitzer:1977},
for each of the scatterings, but one introduces saturation scales.
But, most importantly, these scales are not constants, they depend
on the number of scatterings, and they depend as well on $x^{+}$
and $x^{-}$ \cite{werner:2023-epos4-overview}. In Fig. \ref{saturation-three-pom},
we show for simplicity only gluons, and we do not show the time-like
cascade of further parton emissions from the emitted gluons.

\section{Heavy flavor production in the primary interactions}

\begin{figure}[h]
\noindent \centering{}(a)\hspace*{3cm}(b)\hspace*{3cm}(c)\hspace*{6cm}$\qquad$\\
 \includegraphics[scale=0.25]{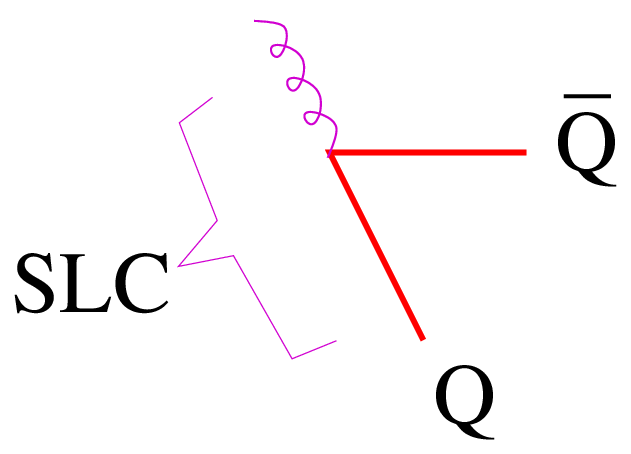}$\qquad$\includegraphics[scale=0.25]{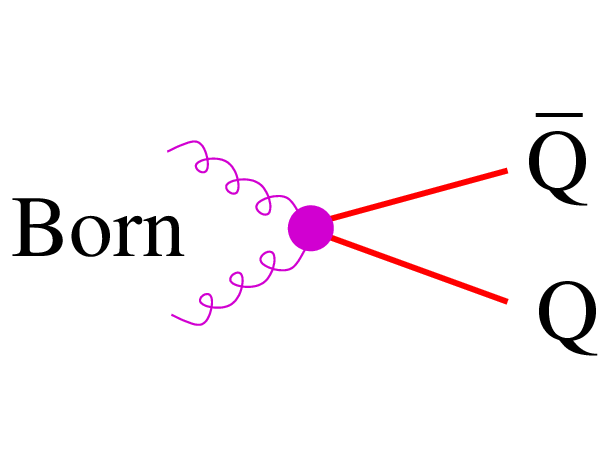}$\qquad$\includegraphics[scale=0.25]{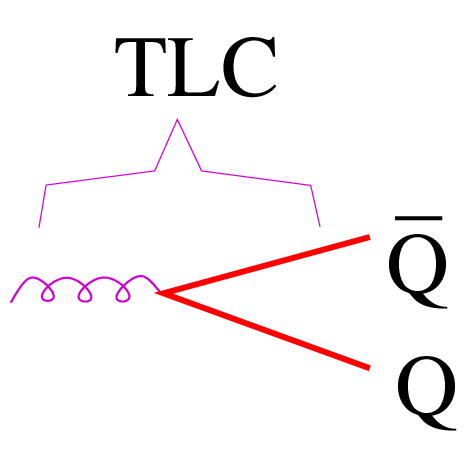}\caption{Different possibilities to create heavy flavor, (a) in the space-like
cascade (SLC), (b) in the Born process, (c) in the time-like cascade
(TLC). \label{charm-3-1}}
\end{figure}

Concerning heavy flavour, we use the general notation of $Q$ for
quarks and $\bar{Q}$ for antiquarks. Heavy flavor may be produced
in different ways, as shown in Fig. \ref{charm-3-1}.
Starting from a gluon, a $Q-\bar{Q}$ pair may be produced in the
space-like cascade, as shown in Fig. \ref{charm-3-1}(a), provided
the virtuality is large enough. The number of allowed flavors is considered
to be depending on the virtuality (variable flavor number scheme).
It is also possible to create a $Q-\bar{Q}$ in the Born process,
via $g+g\to Q+\bar{Q}$ or $q+\bar{q}\to Q+\bar{Q}$ (for light flavor
quarks $q$), as shown in Fig. \ref{charm-3-1}(b), and finally $Q-\bar{Q}$
may be produced in the time-like cascade, via $g\to Q+\bar{Q}$, as
shown in Fig. \ref{charm-3-1}(c).

\begin{figure}[h]
\noindent \centering{}\includegraphics[scale=0.25]{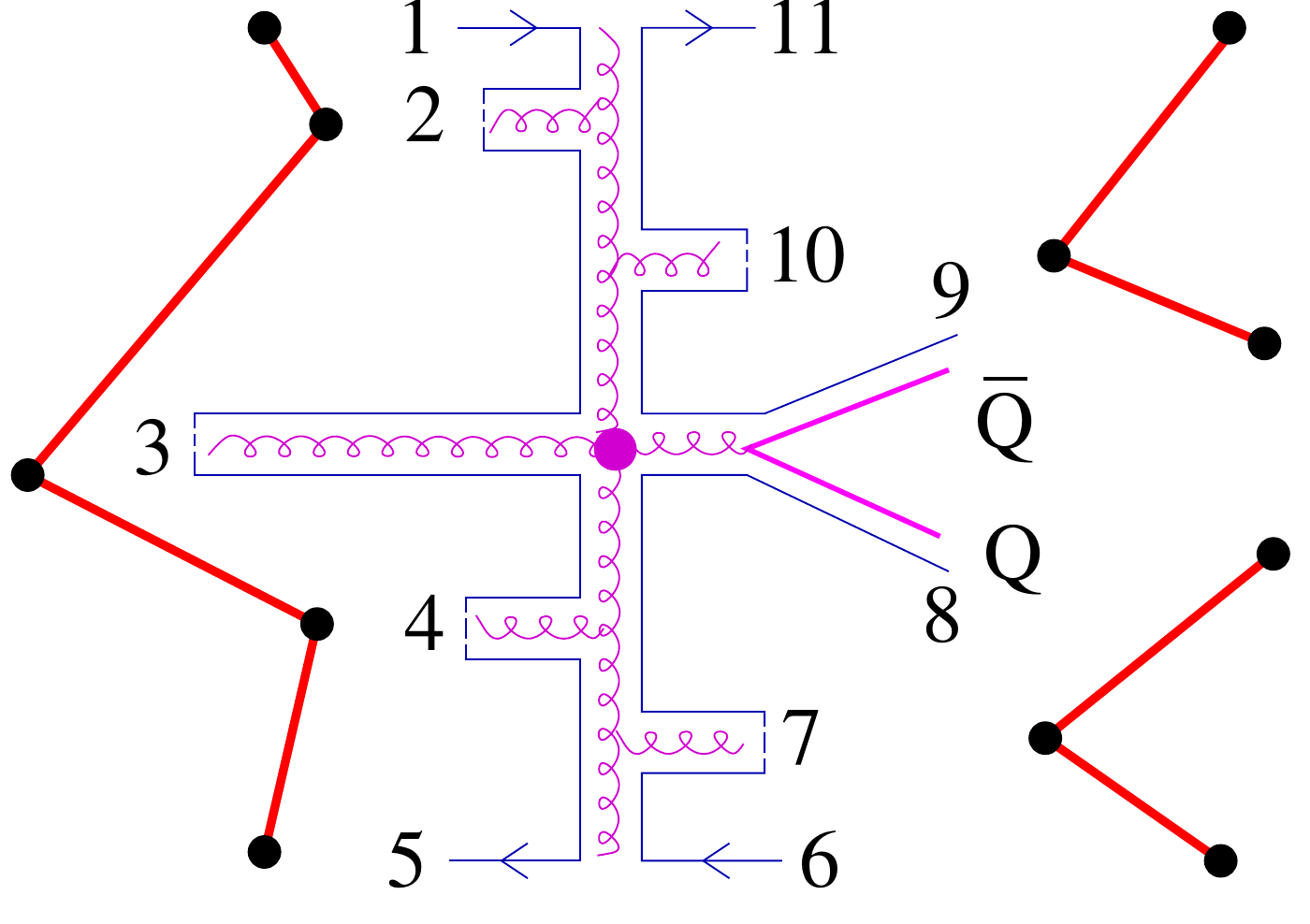}\caption{The chains 1-2-3-4-5, 6-7-8, and 9-10-11 are mapped to kinky strings
(red lines). The black points indicate the kinks, which carry the
parton momenta. \label{charm-4-1-1}}
\end{figure}

Once the parton emissions are done, one considers the corresponding
Feynman diagram, and constructs a color flow picture, which defines
chains of partons by following the color flow, as shown in Fig. \ref{charm-4-1-1}.
These chains of partons are then mapped (in a unique fashion) to kinky
strings, where each parton corresponds to a kink. 

For more details see \cite{werner:2023-epos4-heavy}. The general
mapping procedure (chains of partons to kinky strings) as well as
the string decay procedures into ``string segments'' (which finally
correspond to hadrons) are described in detail in \cite{Drescher:2000ha}.
Now we use the term ``prehadrons'' for the string segments, not
knowing yet if they eventually become hadrons.

\section{EPOS4 core-corona method and fluid evolution }

From the above-mentioned primary interactions, we obtain a more or
less important number of prehadrons. We employ a core-corona procedure
\cite{Werner:2007bf,Werner:2010aa,Werner:2013tya}, where the prehadrons,
considered at a given proper time $\tau_{0}$, are separated into
``core'' and ``corona'' prehadrons, depending on the energy loss
of each prehadron when traversing the ``matter'' composed of all
the others. Corona prehadrons (per definition) can escape, whereas
core prehadrons lose all their energy and constitute what we call
``core'', which acts as an initial condition for a hydrodynamic
evolution \cite{Werner:2013tya,Karpenko_2014}.
\begin{figure}[h]
\begin{centering}
\includegraphics[bb=0bp 0bp 567bp 520bp,clip,scale=0.38]{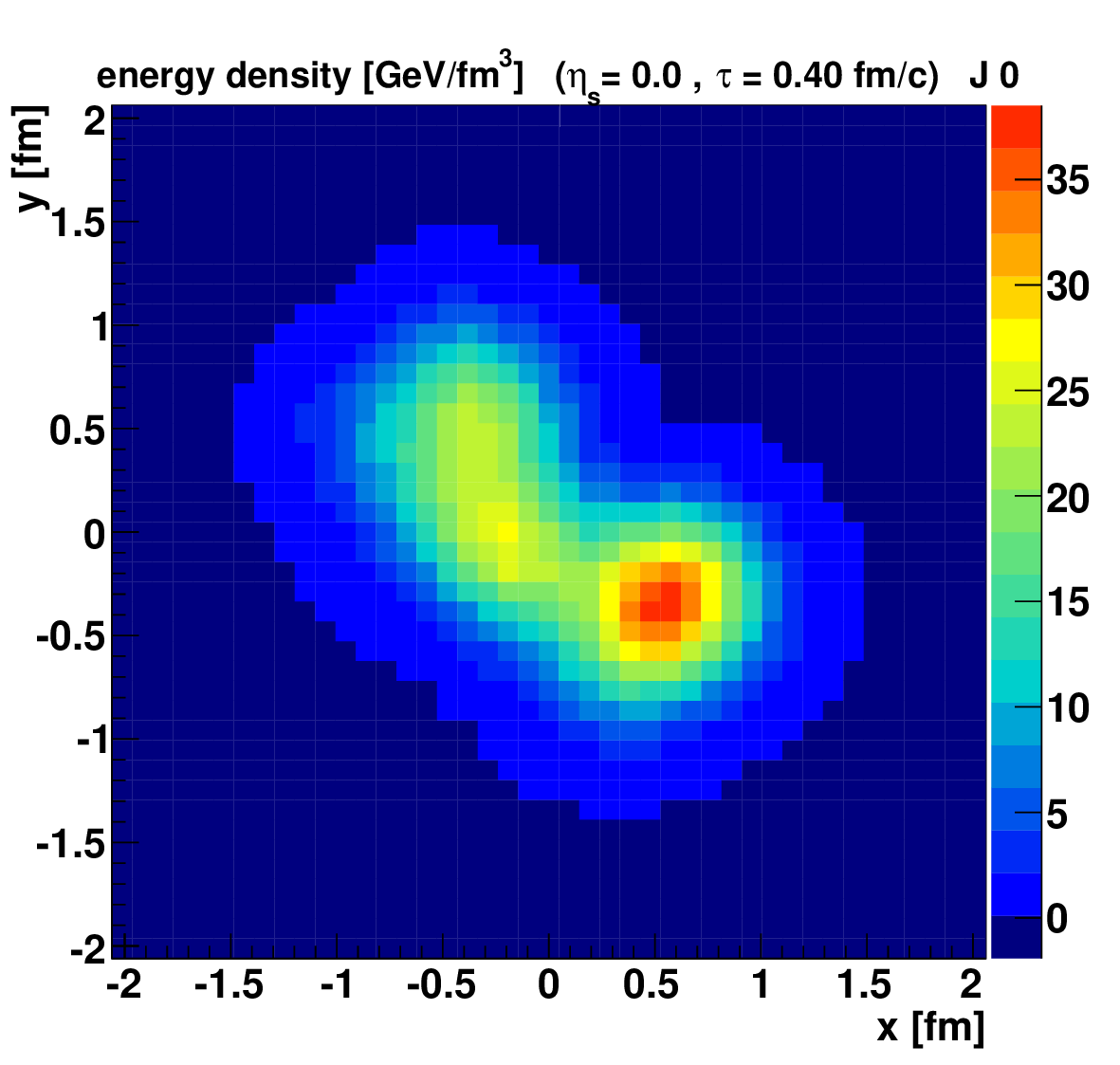} 
\par\end{centering}
\begin{centering}
\includegraphics[bb=0bp 0bp 567bp 520bp,clip,scale=0.38]{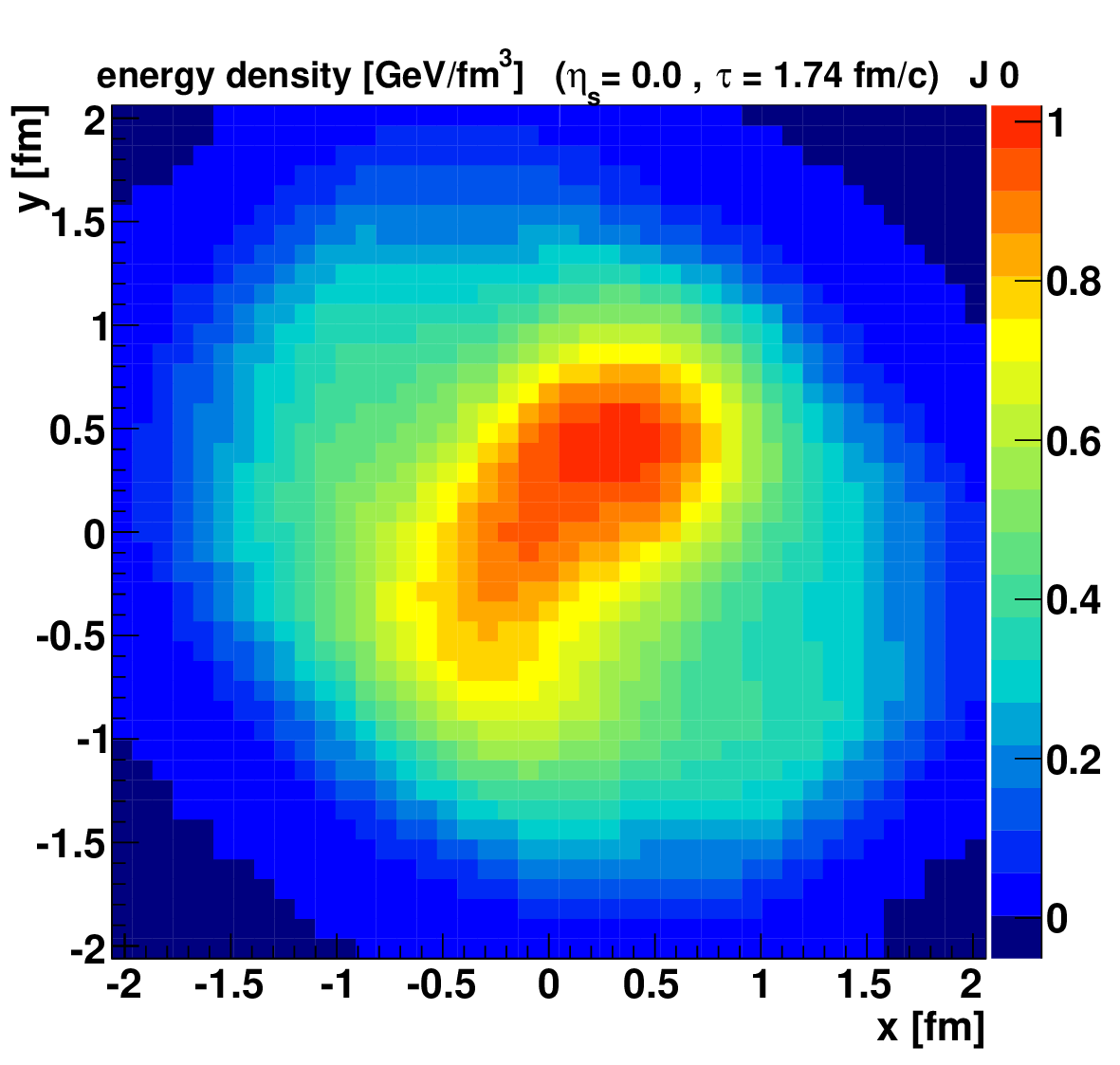}\\
\par\end{centering}
\centering{} \caption{Energy density in the transverse plane ($x,y)$ for proton-proton
scattering involving 6 Pomerons. The upper plot represents the start
time $\tau_{0}$ (of the hydro evolution), and the lower plot a later
time $\tau_{1}$, close to final freeze-out. \label{Energy-density-6-Pomerons}}
\end{figure}
The evolution of the core ends whenever the energy density falls below
some critical value $\epsilon_{\mathrm{FO}}$, which marks the point
where the fluid ``decays'' into hadrons. It is not a switch from
fluid to particles, it is a sudden decay, called ``hadronization''.
Let us consider a (randomly chosen, but typical) 7 TeV proton-proton
scattering event with 6 Pomerons (representing roughly three times
the average). In Fig. \ref{Energy-density-6-Pomerons}, we plot the
energy density in the transverse plane ($x,y)$. We consider two snapshots,
namely at the start time of the hydro evolution $\tau_{0}=0.40\,\mathrm{fm/c}$
(upper plot) and a later time $\tau_{1}$ close to final freeze-out
(lower plot). The initial distribution has an elongated shape (just
by accident, due to the random positions of interacting partons).
One can clearly see that the final distributions are as well elongated,
but perpendicular to the initial ones, as expected in a hydrodynamical
expansion. More examples can be found in \cite{werner:2023-epos4-micro}.

In EPOS4, as discussed in detail in \cite{werner:2023-epos4-micro},
we developed a new procedure of energy-momentum flow through the ``freeze-out
(FO) hypersurface'' defined by $\epsilon_{\mathrm{FO}}$, which allows for
defining an effective invariant mass. It decays according to microcanonical
phase space into hadrons, which are then Lorentz boosted according
to the flow velocities computed at the FO hypersurface. We also developed
new and very efficient methods for the microcanonical procedure \cite{werner:2023-epos4-micro}.
Also in the full scheme, including primary and secondary interactions,
energy-momentum and flavors are conserved.

%\subsection{EPOS4HQ: Heavy quark energy loss in hot medium}
\section{Heavy quark energy loss in EPOS4HQ}

In EPOS4 heavy quarks do not interact and heavy 
hadrons are produced by fragmentation functions,  which describes $e^+e^-$ collisions.
In EPOS4HQ,  if in some spatial regions the energy density of the fluid is above the critical value and therefore a QGP is formed, heavy quarks collide with the partons of the QGP.
We include in this study both, elastic~\cite{Gossiaux:2008jv} and radiative ~\cite{Aichelin:2013mra} collisions.  
To describe these collisions we select in a first step the collision rate, averaging over the thermal distribution of the partons. In a second step we select randomly the momentum of the QGP parton, a light quark or a gluon, from their corresponding thermal distribution assuming that these particles are massless. The thermal distribution is determined by the temperature and the mean velocity at the freeze out surface. The scattering cross sections of the heavy quark with gluons and light quarks are calculated by pQCD matrix elements with a running coupling constant. 

The pQCD elastic scattering cross section diverges for a small momentum transfer in the $t$ and $u$ channels. These infrared divergences are healed by the Debye screening mass $m_D(T)$ of gluons in the hot medium, which is calculated  in the hard thermal loop (HTL) approach.  It serves as a regulator of the propagator of the exchanged gluon.  Scattering at high momentum transfer is, on the contrary,  described by a free gluon propagator. A smooth transition between both regimes can be assured by an effective Debye mass $m_{\rm eff}=\kappa m_D(T)$ in the gluon propagator~\cite{Gossiaux:2008jv}, with $\kappa=0.2$. This effective mass is assumed in our calculations whose details may be found in ref.~\cite{Gossiaux:2008jv}.

The pQCD inelastic scattering cross section has been calculated in~\cite{Aichelin:2013mra}.
This cross section contains 5 matrix elements for gluon emission from the heavy quark  and the light quark and gluon, respectively.  Also for the inelastic cross section the momentum of the plasma particle is chosen by a Monte Carlo approach from the local thermal distribution. 
As in the elastic cross section, the gluon propagator is regulated by $m_{\rm eff}=\kappa m_D(T)$.  For the gluon emission vertex a constant $\alpha_s = 0.3$ is used. The emitted gluon is considered as massless. The different limits of the pQCD cross section calculations as well as more details of the approach have been discussed in Ref.~\cite{Aichelin:2013mra} .

Both, the elastic as well as the inelastic collisions, have been already employed in former EPOS versions to describe heavy meson data in heavy-ion collisions at LHC energies ~\cite{Gossiaux:2009mk,Gossiaux:2010yx}.  We use this theoretical approach to calculate the energy loss of heavy quarks, without modification, also in this new EPOS4HQ version. In this paper the K-factor for elastic as well as for inelastic collisions, which has been varied in the past, Refs.~\cite{Nahrgang:2013saa,Nahrgang:2014vza}, is equal one, so the calculated pQCD cross sections are not modified by an overall factor.

%\subsection{Heavy flavor hadronization in EPOS4HQ}
\section{Heavy flavor hadronization in EPOS4HQ}
When the heavy quark crosses the freeze-out hypersurface, it will be converted into a colorless heavy flavor hadron. This process is non-perturbative and is usually related to soft gluon radiation. 
Whereas in the standard EPOS4 approach (called pure EPOS4 in this study) all heavy quarks hadronize by fragmentation, in EPOS4HQ also hadronization by coalescence may contribute if a QGP is formed. In the coalescence process the heavy quark combines with one/two nearby fluid (QGP) partons to form final-state mesons or baryons. Heavy quarks, which do not traverse a QGP, create hadrons by fragmentation.   

In the coalescence model, the heavy quarks coalesce with light quarks to a hadron $m$ when the charm passes the hadronization hypersurface determined by the critical energy density. The light (anti) quarks are assumed to be thermalized.
The differential yield of the heavy hadron is given by:
\begin{eqnarray}
{dN\over d^3{\bf P}}&=&g_H\sum_{N_c} \int \prod_{i=1}^k{d^3{\bf p}_i\over (2\pi)^3} f_i({\bf p}_i)W_m({\bf p}_1,..,{\bf p}_i)\nonumber\\
&\times& \delta^{(3)}\left({\bf P}-\sum_{i=1}^k{\bf p}_i\right) ,
\label{eq.coal}
\end{eqnarray}  
where $g_H$ is the degeneracy factor of color and spin. $k=2(3)$ for mesons (baryons). ${\bf P}$ and ${\bf p}_i$ are the momenta of heavy flavor hadron and the constituent quarks, respectively. The delta function conserves the momentum.  The summation is performed over all heavy quarks in the system.

$f_1({\bf p}_1)=(2\pi)^3\delta^{(3)}\left({\bf p}_c-{\bf p}_1\right)$ is the normalized momentum space distribution of the heavy quark and  $f_i({\bf p}_i)$ for $i>1$ is the momentum space distribution of the constituent quarks in the heavy hadron.  
$W_m({\bf p}_1,..,{\bf p}_i)$ is the Wigner density of the heavy hadron $m$ in momentum space, which can be constructed from the hadron wave function. The hadron wave function can be approximated by a three-dimensional harmonic oscillator state with the same root mean square radius. For the ground state of charmed meson, the Wigner density in the center-of-mass (CM) frame can be expressed as,
%, and is  for the two-body case defined as
%\begin{eqnarray}
%W({\bf r}, {\bf p}_r)=\int d^3{\bf y}e^{-i{\bf p}_r\cdot {\bf y}}\psi({\bf r}+{{\bf y}\over 2})\psi^*({\bf r}-{{\bf %y}\over 2}),
%\end{eqnarray}  
%where ${\bf r}$ and ${\bf p}_r$ are relative distance and momentum, respectively. The relative wave function $\psi$ %can be obtained by solving the two-body Schr\"odinger equation.  In our study, the coalescence probability is given %by the momentum space Wigner function, which can be obtained by integrated over the spatial space and expressed ($S$-wave),
\begin{eqnarray}
W(p_r)&=&{(2\sqrt{\pi}\sigma)^3}e^{-\sigma^2{p_r}^2},
\end{eqnarray} 
% The integral will be canceled by the volume factor associated with the momentum distribution functions of light quarks; 
where $p_r$ is the relative momentum between  two constituent quarks in the CM frame. It's normalized, $\int W(p_r)d^3{\bf p}_r/(2\pi)^3=1$. We use $
 p_r={|E_2{\bf p}_1-E_1{\bf p}_2|/( E_1+E_2)}$, where  $E_1 ({\bf p}_1)$ and $E_2({\bf p}_2)$ are the energies and momenta of the quark or antiquark in the CM frame, respectively. The width $\sigma$ in the Wigner density is controlled by the root-mean-radius as shown next.  
Baryons are treated as two two-body systems (baryons are produced by recombining two quarks first and then using their center of mass to recombine with the third one.). 
%---------------------------------------------------------------------
\begin{figure}[!htb]
\includegraphics[width=0.35\textwidth]{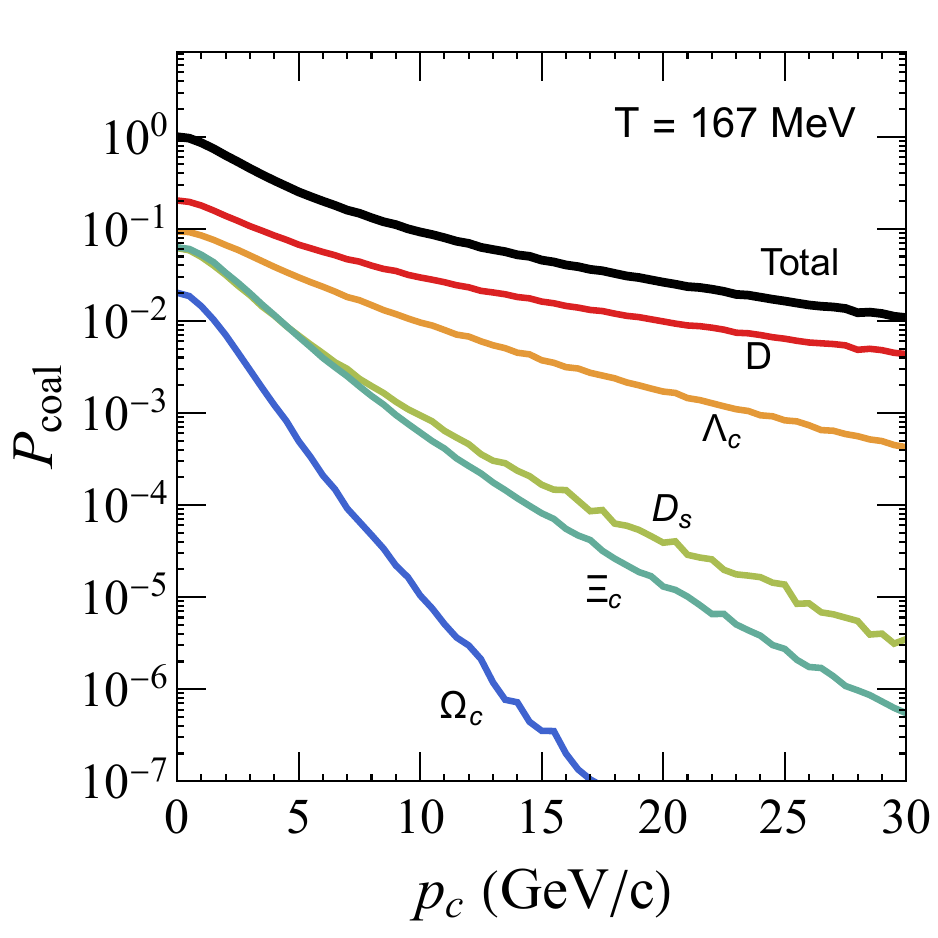}
\caption{Momentum dependent coalescence probability of a charm quark into various charmed hadrons in a hot and static medium with temperature $T_{\rm FO}=167\rm MeV$. Here $D$ is the sum of $D^0$ and $D^+$. $\Xi_c$ is the sum of $\Xi^0$ and $\Xi^+$.}
\label{fig_coal_prob}
\end{figure}
%---------------------------------------------------------------------

The light quarks are assumed to be thermalized. 
The quark masses are $m_{u/d}=0.1\rm GeV$, $m_{s}=0.3\rm GeV$, and $m_{c}=1.5\rm GeV$. 
The heavy quark coalescence probability in a static hot medium with a temperature $T_{\rm FO}=167 \rm MeV$ corresponding to a critical energy density of $\epsilon_{\rm FO}=0.57\rm GeV/fm^3$
is obtained by integrating Eq.~\ref{eq.coal}.
The root-mean-square radius of  a heavy hadron is defined as $\langle r^2\rangle=\sum_{i=1}^k \langle ({\bf r}_i-{\bf R})^2\rangle$ with the quark coordinate ${\bf r}_i$ and the CM coordinate ${\bf R}$. For charmed mesons, $\langle r^2\rangle={3\over 2}{m_c^2+m_q^2\over (m_c+m_q)^2}\sigma^2$. The root-mean-square radius of the ground state charmed meson can be calculated by the two-body Dirac equation~\cite{Zhao:2018jlw}. It gives $\sqrt{\langle r^2\rangle}\approx0.85\rm fm$, for in-medium $D^0$.  This leads to a width  $\sigma=3.725\rm GeV^{-1}$. We take the same width $\sigma$ for $D_s$ and any two-quark system in charmed baryons. 

The coalescence probability of excited states, which can strongly decay into the ground states, is estimated via the statistic model. There the hadron density at the temperature $T_{\rm FO}$ is given by~\cite{Andronic:2007zu},
$n_i={g_i\over2\pi^2}T_{\rm FO}m_i^2K_2 ({m_i\over T_{\rm FO}})$. $g_i$ is the spin isospin degeneracy. $m_i$ is the mass of the hadron. $K_2$ is the second-order Bessel function. 
In our study, we consider almost all possible excited states, also the missing baryons, which are predicted by the quark model~\cite{Ebert:2011kk} and lattice QCD~\cite{Bazavov:2014yba,Padmanath:2014lvr}.  For each 
ground state hadron $D$, $D_s$, $\Lambda_c$, $\Xi_c$, and $\Omega_c$  we calculate the density of each excited state $m$ and define the momentum-independent ratio $R^m=n_{\rm excited}^m/n_{\rm ground}$. Finally we sum over all excited states $R=\sum R^m$ and multiply the ground state $p_T-$dependent coalescence probability by $1+R$ to obtain the coalescence probability for prompt charmed hadrons as shown in Fig. \ref{fig_coal_prob}. The sum of the effective $p_T$ distributions for all hadrons gives the total coalescence probability $P_{\rm coal}^{\rm Total}(p_T)$ as shown with the black line in Fig. \ref{fig_coal_prob}.
Heavy quarks, which do not hadronize via coalescence, will fragment into a heavy-flavor hadron. The fragmentation probability is therefore $1-P_{\rm coal}^{\rm Totel}$.
The fragmentation function we employ are those from the heavy quark effective theory~\cite{Braaten:1994bz,Cacciari:2005rk} and the fragmentation ratios to various charmed hadrons are taken as the $e^+e^-$ collisions~\cite{Lisovyi:2015uqa}. The evolution of heavy flavor hadrons in the hadronic phase is controlled by the UrQMD transport model but generally hadronic rescattering is negligible, as we will show.
%\jx{why it is negligible? this should be one of the conclusions of our results.}

%==================================
\section{Results and analysis}

When in $pp$ collisions a QGP is created because locally the energy density exceeds the critical energy density, there are two ways in which heavy hadrons are formed: by fragmentation and by coalescence, as described in the last section. It is the coalescence contribution that changes the observables in medium and high multiplicity $pp$ collisions as compared to those at low multiplicity and observed in $e^+e^-$ events.
We focus in the following on four observables, the transverse momentum spectra, yield ratios, elliptic flow, and correlations.
%-------------------------------
\begin{figure}[!htb]
$$\includegraphics[width=0.47\textwidth]{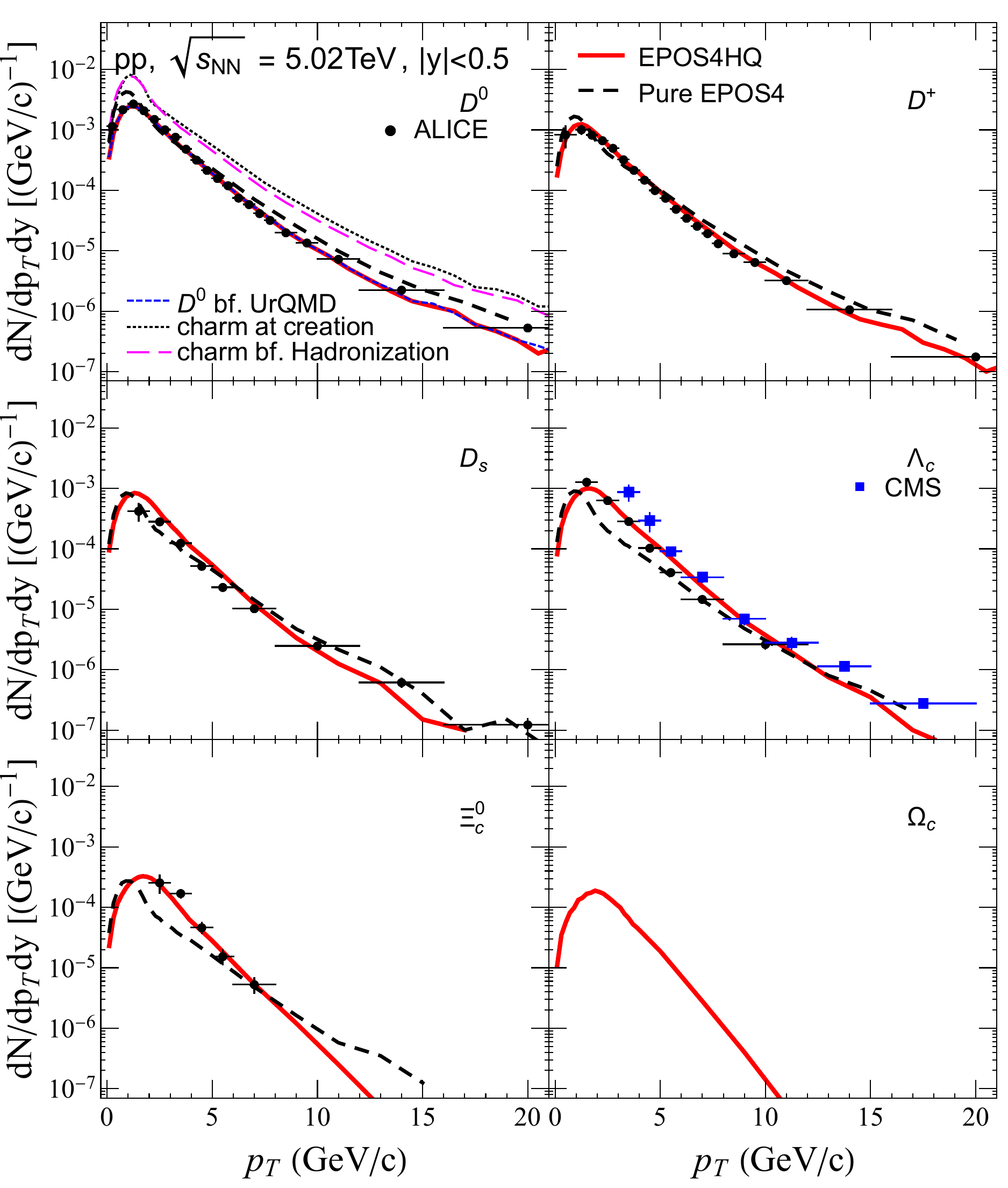}$$
\caption{$p_T$ spectrum of $D^0$, $D^+$, $D_s$, $\Lambda_c$, $\Xi_c^0$, and $\Omega_c$ in 5.02TeV $pp$ collisions. The experiment data are from the ALICE~\cite{ALICE:2020wfu,ALICE:2021psx} and CMS~\cite{CMS:2023frs}.}
\label{fig_spect}
\end{figure}
%---------------------------------------------------------------------

{\bf Transverse momentum spectrum}:  The EPOS4HQ transverse momentum spectra 
of $D^0$, $D^+$, $D_s$, $\Lambda_c$, $\Xi_c^0$, and $\Omega_c$  hadrons are shown in  Fig.~\ref{fig_spect} as red line and compared to the available experimental data  from ALICE~\cite{ALICE:2020wfu,ALICE:2021psx} and CMS~\cite{CMS:2023frs}. For $D^0$  mesons we display as well the distribution of the charm quarks at creation (dotted black line) and before hadronization (dashed magenta line), as well that of $D^0$ mesons immediately after hadronization, before the hadronic rescattering (dashed blue line), which is almost identical to the red line. The black-dashed lines display the results from the pure EPOS4 with fragmentation only. A significant difference can be observed between the pure EPOS4 and EPOS4HQ in the charmed bayons spectra. This reveals the importance of the coalescence in high energy $pp$ collisions. We see that the energy loss of the charm quarks in the QGP change the $p_T$ spectrum considerably. The difference in the $p_T$ spectrum of charm quarks at creation and the $D^0$ mesons observed in the detectors is almost exclusively due to the hadronization. Consequently, the transverse momentum spectra are not very sensitive to the presence of a QGP. We observe furthermore that our results reproduce quite well the experimental data.
%---------------------------------------------------------------------
\begin{figure}[!htb]
$$\includegraphics[width=0.23\textwidth]{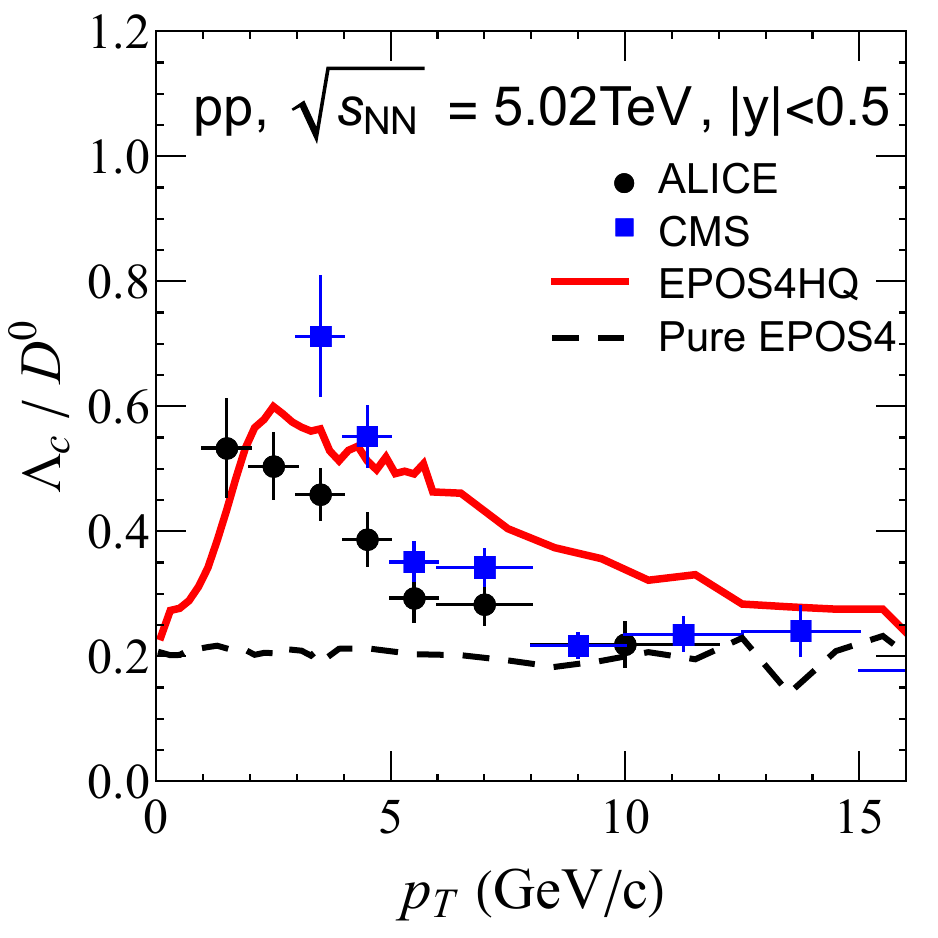}\includegraphics[width=0.23\textwidth]{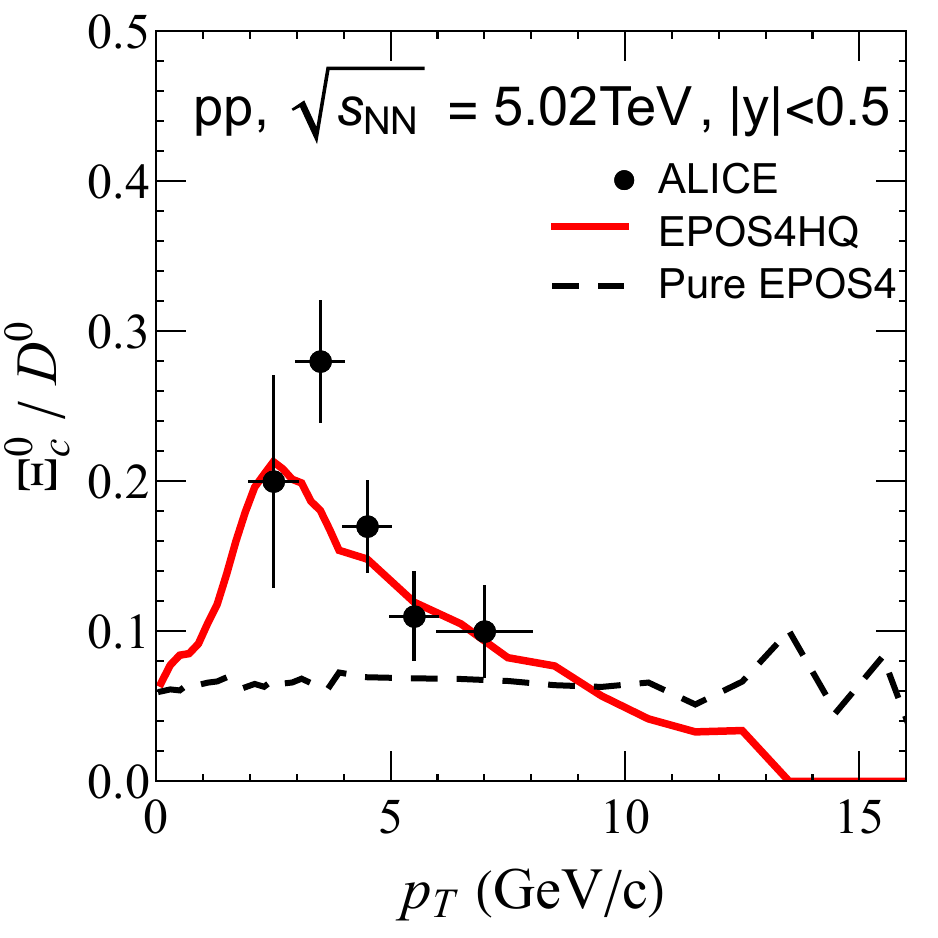}$$
\caption{Yield ratio of $\Lambda_c/D^0$ and $\Xi_c^0/D^0$ as a function of  the transverse momentum $p_T$ in 5.02TeV $pp$ collisions. The experiment data are from the ALICE~\cite{ALICE:2020wfu,ALICE:2021psx} and CMS~\cite{CMS:2023frs}.}
\label{fig_yield}
\end{figure}
%---------------------------------------------------------------------

\begin{figure}[!htb]
$$\includegraphics[width=0.35\textwidth]{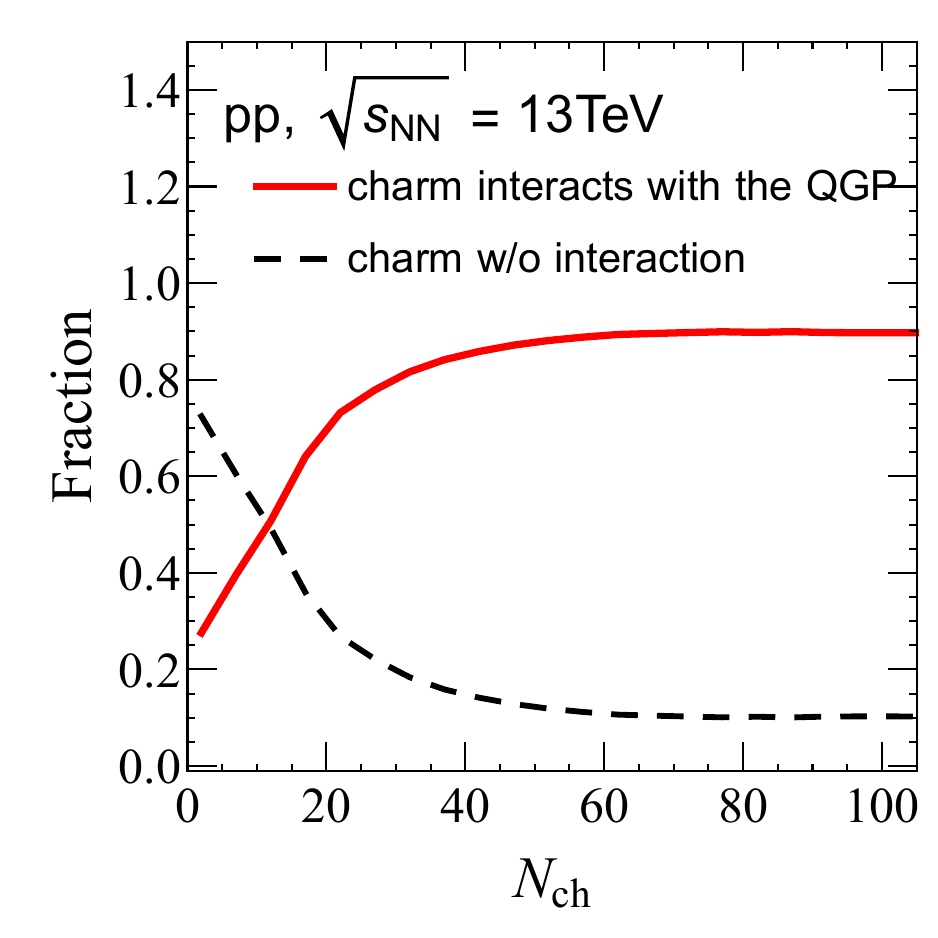}
$$
\caption{ Fraction of charm quarks which interact with the QGP in $pp$ collisions at 13 TeV as a function of the multiplicity of charged particles with $|\eta|<1$.}
\label{fig_charm_nch}
\end{figure}

{\bf Yield ratio:} The measured yield ratios of charmed baryons and mesons, $\Lambda_c/D^0$ and  $\Xi_c^0/D^0$,  in $pp$ collisions~\cite{CMS:2010ifv},  are shown in  Fig.~\ref{fig_yield}. We compare pure EPOS4 and EPOS4HQ results with the experimental data.  EPOS4HQ shows  a strong enhancement of this ratio at low $p_T$ and describes the experimental data quite reasonably. Pure EPOS4, in which  hadrons are exclusively produced via fragmentation, shows a $p_T$ independent ratio.   This indicates that the origin of the enhancement by roughly a factor of 2.5  is the rescattering of heavy quarks with QGP partons and the subsequent hadronization via coalescence. 
The difference of this ratio in $e^+e^-$ and $pp$ collisions has been interpreted as a sign of process dependent fragmentation functions~\cite{Christiansen:2015yqa}. Our calculation does not point into this direction.
The origin is  rather a more complex reaction scenario in pp as compared to $e^+e^-$,  in particular the creation of  regions with a high energy density what can be interpreted as the creation of a QGP.
%---------------------------------------------------------------------
\begin{figure}[!htb]
$$\includegraphics[width=0.23\textwidth]{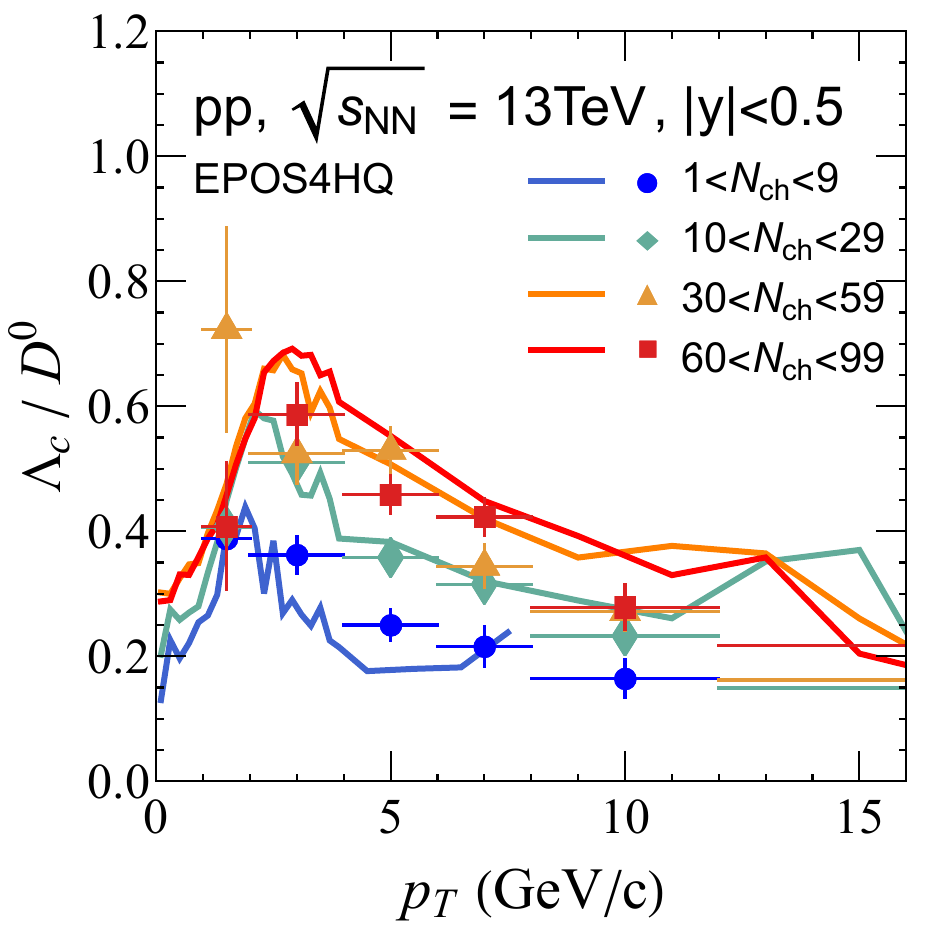}\includegraphics[width=0.23\textwidth]{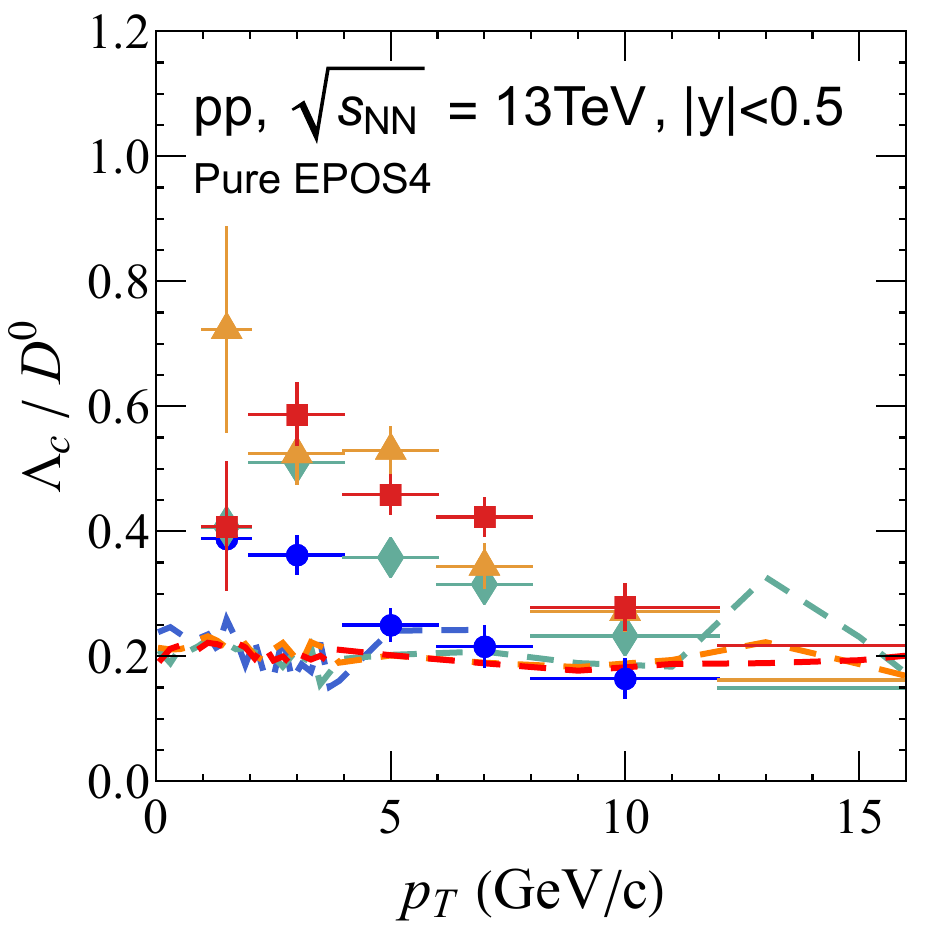}$$
\caption{ $\Lambda_c/D^0$ ratio as a function of $p_T$ in 13TeV $pp$ collisions. The results are shown with the EPOS4HQ (left panel) and pure EPOS4 (right panel) in different multiplicity range. The experiment data are from the ALICE~\cite{ALICE:2021npz}. Here the multiplicity bin is selected in midrapidity $|\eta|<1$.}
\label{fig_lc_D0_nch_all}
\end{figure}
%---------------------------------------------------------------------
In Fig.~\ref{fig_lc_D0_nch_all} we study the momentum dependence of $\Lambda_c/D^0$ in four multiplicity bins,
left for EPOS4HQ and right for pure EPOS4. In pure EPOS4 this ratio is, as expected,  constant and independent of the multiplicity. In EPOS4HQ this ratio is considerably higher and approaches only for decreasing multiplicity the value of EPOS4. With decreasing multiplicity, the energy density of the produced partons gets lower. As a result, both the size of the QGP and the charm quark fraction in the QGP are decreasing. In Fig.~\ref{fig_charm_nch} we display the multiplicity-dependent charm fraction. The charm quarks, which do not pass a QGP, do not interact and hadronize only via fragmentation, while those which pass the QGP will interact with the thermal partons and hadronize finally via coalescence plus fragmentation. The increase of the fraction of charm quarks, which pass a QGP, leads to an enhancement of the coalescence contribution and as a result of the $\Lambda_c/D^0$ ratio.
%---------------------------------------------------------------------
%---------------------------------------------------------------------

\begin{figure}[!htb]
$$\includegraphics[width=0.35\textwidth]{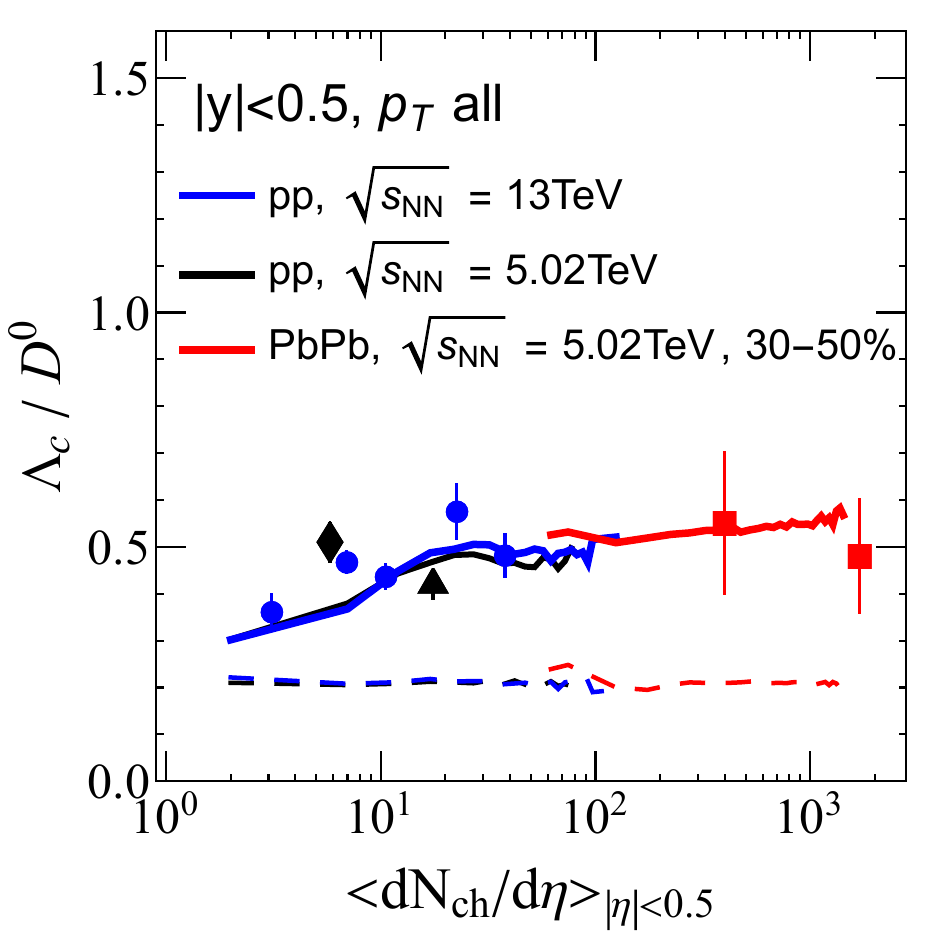}
$$
\caption{ Multiplicity dependent of $\Lambda_c/D^0$. The results are shown for the EPOS4HQ (solid lines) and pure EPOS4 (dashed lines) in 5.02 TeV, 13 TeV $pp$ and 30-50\% central 5.02 TeV $PbPb$ collisions. The experimental data is from Ref.~\cite{ALICE:2021npz}.}
\label{fig_ratio_v2_nch}
\end{figure}

Fig~\ref{fig_ratio_v2_nch} shows the $p_T$ integrated ratio $\Lambda_c/D^0$ as a function of the multiplicity in comparison with experimental data~\cite{ALICE:2021npz}. This enhancement increases smoothly from low (where the results agree with those measured in $e^+e^-$ collisions) to high multiplicity  and finally reaches saturation as the fraction of the charm quarks, which pass a QGP (Fig. \ref{fig_charm_nch}).

%---------------------------------------------------------------------
%---------------------------------------------------------------------

\begin{figure}[!htb]
\includegraphics[width=0.35\textwidth]{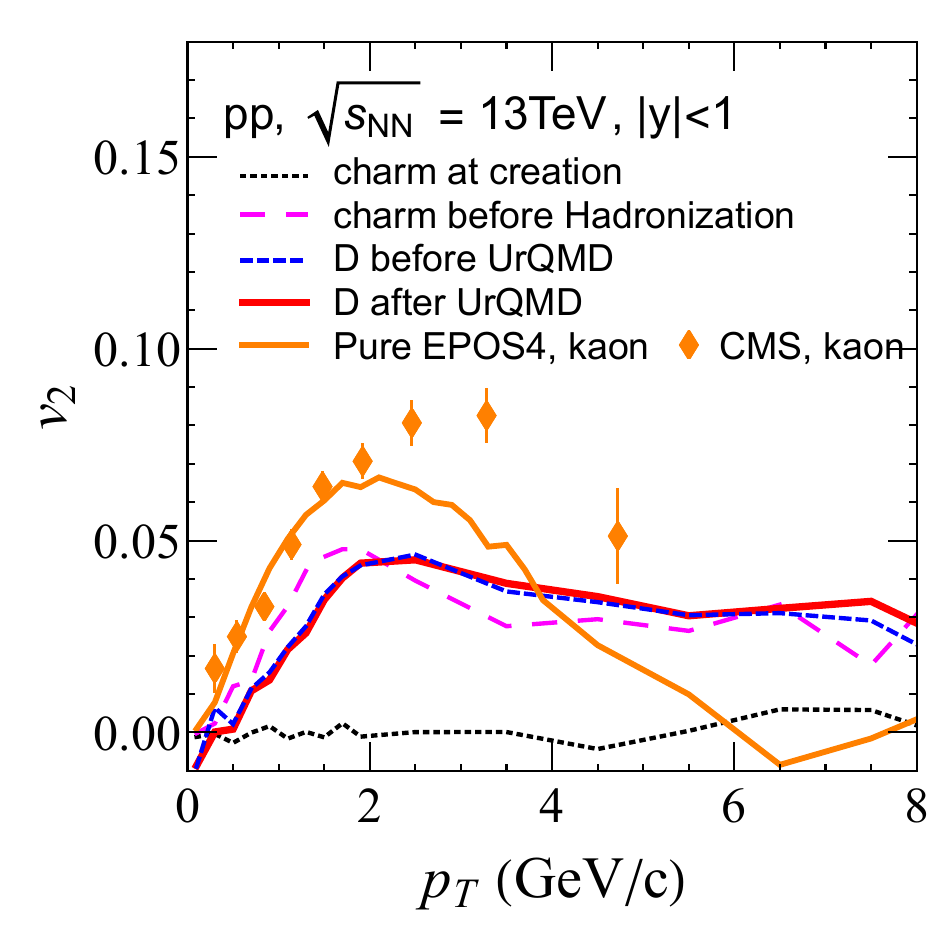}
\caption{Elliptic flow of charm quarks at creation and after passing through the QGP
as well as that of $D$ mesons immediately after production and after hadronic rescattering
using the UrQMD model. $v_2$ of the $K$ meson in comparison with the experimental data ($v_2\{2\}$) from the CMS~\cite{CMS:2016fnw} is displayed as well.}
\label{fig_v2pt_all}
\end{figure}

{\bf Elliptic flow $v_2$ of $D^0$ meson}:  Another observable, which is sensitive to collectivity,
is the elliptic flow $v_2$.  It is displayed, as a function of $p_T$, in Fig.~\ref{fig_v2pt_all}.
Initially heavy quarks show an isotropic distribution 
%---------------------------------------------------------------------
%---------------------------------------------------------------------
%---------------------------------------------------------------------
\begin{figure}[!htb]
$$\includegraphics[width=0.23\textwidth]{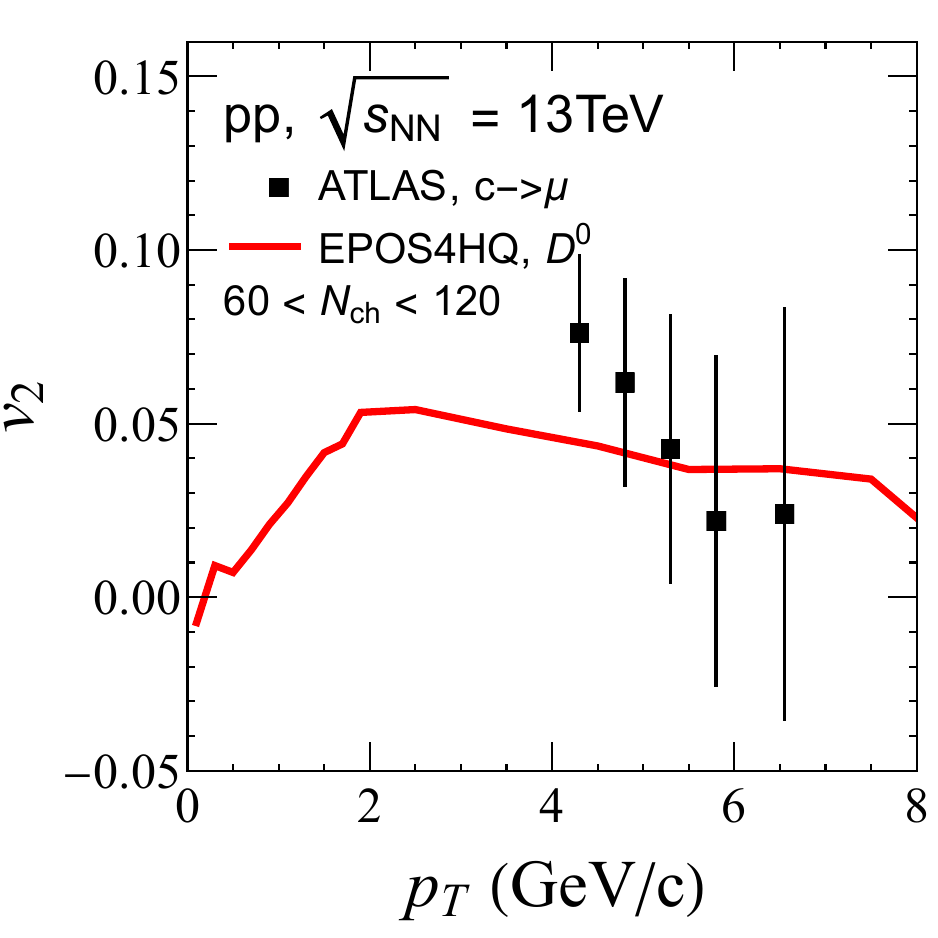}\includegraphics[width=0.23\textwidth]{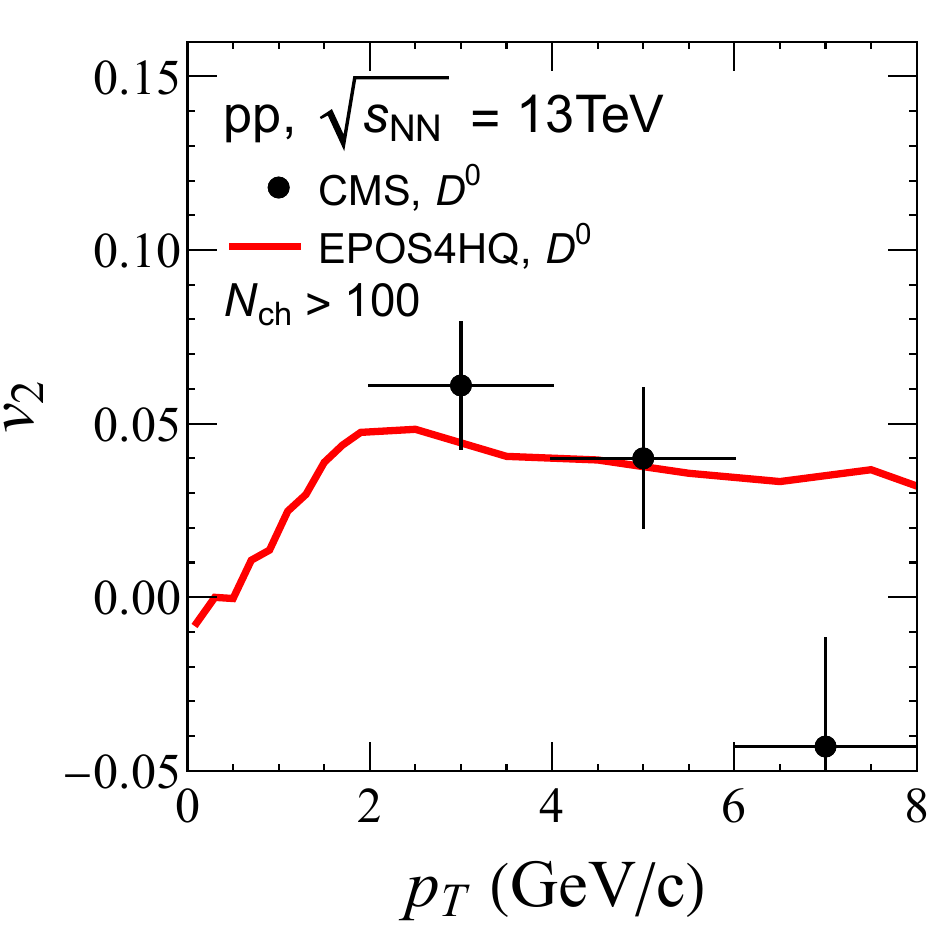}$$
\caption{$D^0$ meson $v_2$ as a function of transvers momentum $p_T$ in 13TeV $pp$ collisions. The experiment data are from the ATLAS~\cite{ATLAS:2019xqc} and CMS~\cite{CMS:2020qul}.}
\label{fig_v2pt_D}
\end{figure}
%---------------------------------------------------------------------
(black dotted line).  They gain $v_2$ by interactions with the QGP partons before hadronization (dashed magenta line). $v_2$ of $D$ mesons immediately after hadronization is shown as short dashed blue line, whereas the result after hadronic rescattering, employing the UrQMD model, is shown as a red line. It is evident that the $v_2$ of heavy mesons is created while the $c$-quark passes the QGP. Hadronization and hadronic rescatter modify $v_2$ only marginally. We plot as well the $v_2\{4\}$ of kaons and compare the result to $v_2 \{2\}$ CMS data to show that $v_2$ in the light sector is described by EPOS4.
%---------------------------------------------------------------------
\begin{figure}[!htb]
\includegraphics[width=0.35\textwidth]{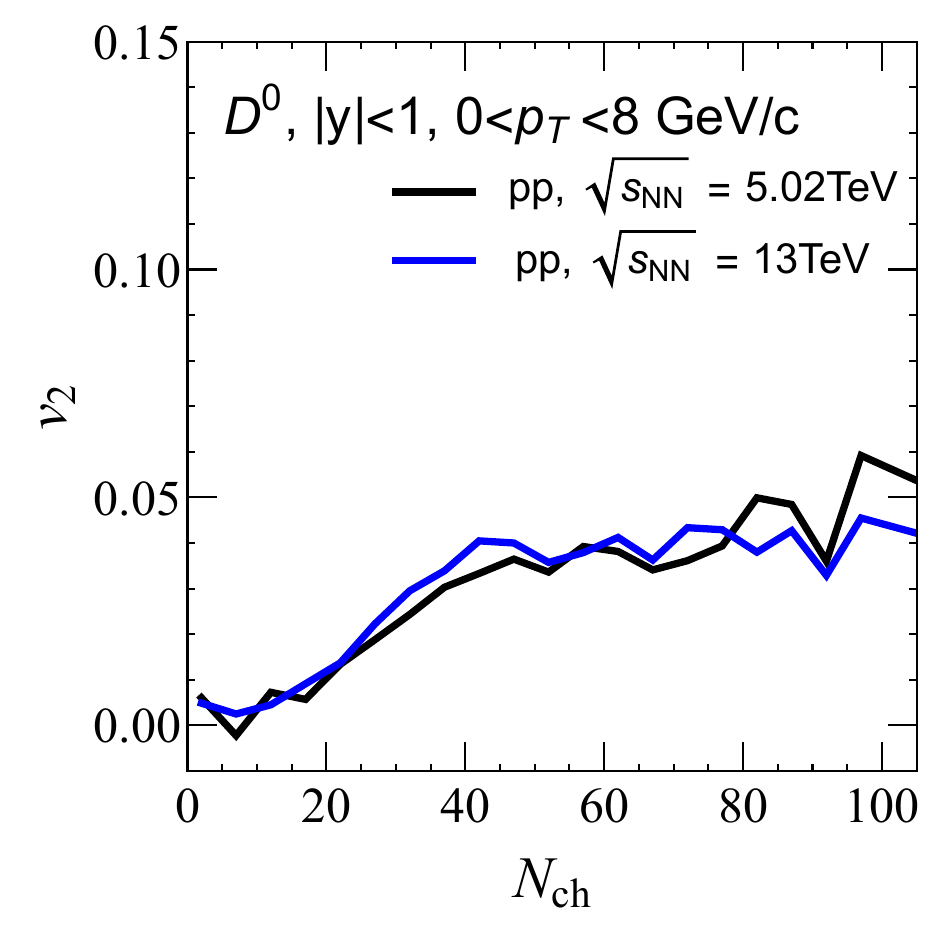}
\caption{Dependence of $v_2$ of $D^0$ mesons on the charged particle multiplicity with $|\eta|<1$.  . The EPOS4HQ results are shown for $\sqrt{s_{\rm NN}}$=5.02 TeV and 13 TeV. }
\label{fig_v2_nch}
\end{figure}
%---------------------------------------------------------------------
%\kw{describe this figure}
To be sure that this is not an artifact due to a too high $v_2$ value of the QGP partons, we compare in Fig.~\ref{fig_v2pt_all} as well $v_2(p_T)$ of kaons (orange line) with experiment and find that our calculation agrees with the experimental finding at least at low $p_T$. Hence the $v_2$ of the QGP partons is correctly reproduced. We note that $v_2(p_T)$ of $D$-mesons is well below the $v_2(p_T)$ of the kaons. In pure EPOS4 calculations, where heavy hadrons are produced by fragmentation, $v_2$ of heavy hadrons is zero, as expected. 

Selecting for the EPOS4HQ results the same multiplicity bins as in the experiments, we can compare our calculations with the experimental data of the ATLAS and CMS collaboration. This comparison is shown in  Fig.~\ref{fig_v2pt_D}. The calculated (as compared to the kaons lower) $v_2$ values (see Fig.~\ref{fig_v2pt_all} ) agree quantitatively with the experimental data for different multiplicities as measured by the two collaborations.

The multiplicity  dependence of $v_2$ is shown in Fig. \ref{fig_v2_nch}.  Also here one can see that $v_2$ increases with multiplicity. As expected from Fig. \ref{fig_charm_nch} we reach an asymptotic value at $N_{ch}\approx 50$.

%Due to the dominance of the t-channel elastic scattering, we can obtain a finite $v_2$, even if the energy loss of the heavy quarks is small, as shown in Fig~\ref{fig_spect} where the black line is charm quark distribution at creation while blue dashed line is that before the hadronization.
%{\bf Elliptiv flow $v_2$}
\begin{figure}[!htb]
\includegraphics[width=0.35\textwidth]{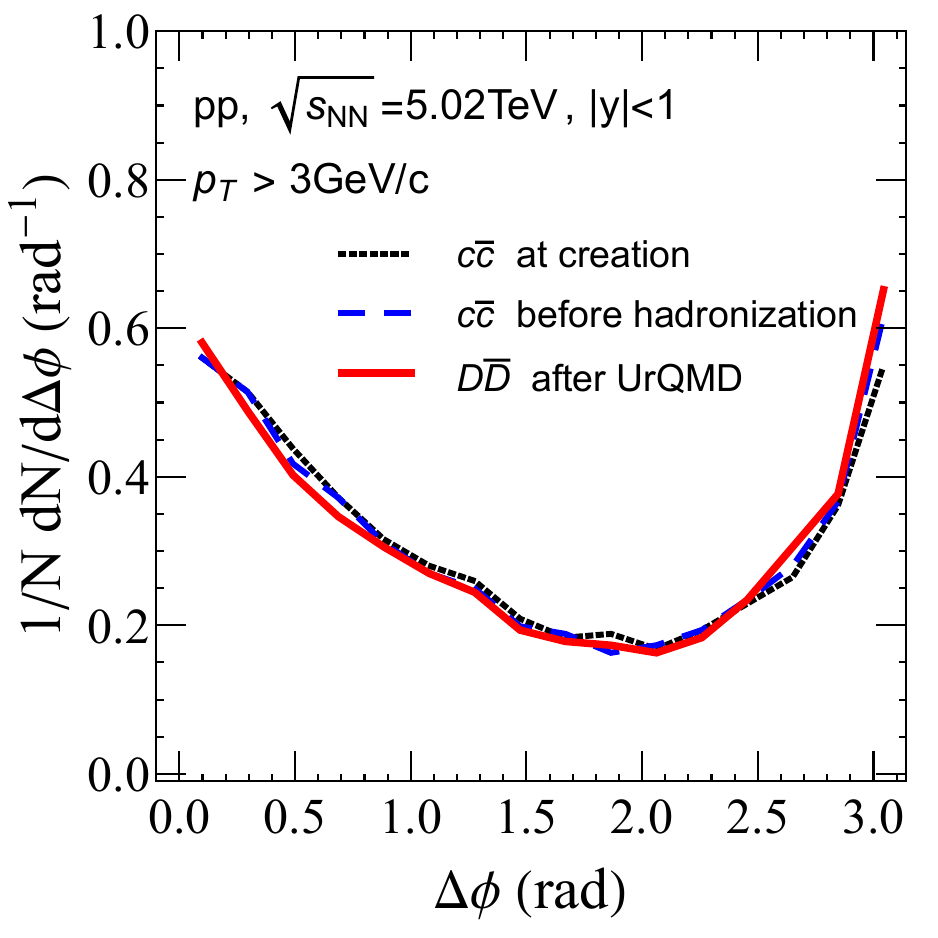}
\caption{ $c\bar c$ and $D\bar D$ correlations in 5.02TeV $pp$ collisions. A selection of $p_T>3$ GeV/c for both trigger and associated particles is used in all cases.}
\label{fig_DDbar_corr}
\end{figure}

{\bf Correlations}:
Pure EPOS4 reproduces well $DD$ and $D\bar D$ correlations ~\cite{Werner:2023fne}.  The correlations of $c\bar c$ and $D\bar D$ in EPOS4HQ are displayed in Fig.~\ref{fig_DDbar_corr} for $p_T>3$ GeV (for both heavy quarks or mesons) and are more pronounced with this $p_T$ cut than without.
Neither the interaction of heavy quarks with QGP partons (blue-dashed line) nor the subsequent hadronic rescattering (red line) has an influence on these correlations and therefore EPOS4 and the EPOS4HQ give almost same result, which agrees nicely with experiment ~\cite{Werner:2023fne}. So we have to conclude that the creation of a QGP does not affect the $D\bar D$ correlations for $p_T>3$ GeV . This gives the chance to probe the initial charm production via the final correlation of charmed hadron.

%\pbg{How good is it compatible with the result on $v_2$ ? Is the result the same if one makes 2 classes for the azimuthal correlation (one along the event plane and one $\perp$ to the event plane} 

\section{Discussion and Conclusion } 

We have investigated particle spectra, baryon to meson ratios, the elliptic flow $v_2$, correlations of heavy mesons, employing the new EPOS4HQ approach, which is compatible with the experimental light hadron observables, in $pp$ as well as in heavy-ion collisions at LHC  and RHIC energies. Employing a system size independent critical energy density for the formation of a QGP we observe  that also in a $pp$ collisions a QGP can be formed. 

If a QGP is formed the heavy quarks interact with the QGP constituents. These interactions  are the origin of the finite $v_2$ values. Our result agree with the experimental findings. The hadronization itself and the later hadronic rescattering change the elliptic flow of D-mesons only little. Due to the dominance of the t-channel elastic scattering, we can obtain a finite $v_2$, even if the energy loss of the heavy quarks in the QGP is small and the change of the transverse momentum spectrum only marginal.
The difference between the $p_T$ spectrum of heavy quarks at production and the $p_T$ spectrum of the observed $D$ mesons is almost exclusively due to the hadronization process.

In EPOS4 hadrons are formed by string fragmentation. The hadronization of heavy quarks, which pass a QGP, is described in EPOS4HQ in addition by coalescence. We observe that coalescence enhances strongly the heavy baryon yield at low $p_T$ reproducing the observed experimental enhancement as compared to $e^+e^-$ collisions.  That the heavy baryon to heavy meson ratio in EPOS4HQ increases with the charged particle multiplicity for low multiplicities and saturates at higher multiplicities, in accordance with the experimental findings, presents further evidence that the formation of a QGP is at the origin of this enhancement.

We can therefore conclude that those observables, which are sensitive to the formation of a QGP, suggest that $pp$ collisions are not elementary reactions but the low system size limit of heavy-ion reactions in which the formation of a QGP is observed if the energy density exceeds a system size independent critical value.

\vspace{1cm}
\noindent {\bf Acknowledgement}: This work is funded by the European Union’s Horizon 2020 research and innovation program under grant agreement No. 824093 (STRONG-2020).

%=====================
\bibliographystyle{apsrev4-1.bst}
\bibliography{Ref}

\end{document}